\documentclass[10pt]{amsart}
\usepackage{macrosgensym}
\usepackage{fullpage}

\usepackage{epigraph}
\usepackage{comment}
\usepackage{parskip}
\def\op{{\rm op}}
\DeclareMathOperator{\Map}{{\rm Map}}

\DeclareMathOperator{\Mfld}{\mathsf{Mfld}}
\DeclareMathOperator{\AbGrp}{\mathsf{AbGrp}}
\DeclareMathOperator{\Top}{\mathsf{Top}}
\DeclareMathOperator{\Open}{\mathsf{Open}}
\DeclareMathOperator{\Disk}{\mathsf{Disk}}

\DeclareMathOperator{\Hom}{Hom}

\DeclareMathOperator{\Spaces}{\cS {\sf paces}}
\DeclareMathOperator{\Alg}{\cA {\sf lg}}
\DeclareMathOperator{\Shv}{\cS {\sf hv}}
\DeclareMathOperator{\dStk}{\cS {\sf tk}}
\DeclareMathOperator{\Sol}{Sol}
\DeclareMathOperator{\CAlg}{\cC {\sf Alg}}
\DeclareMathOperator{\Conn}{Conn}
\DeclareMathOperator{\Flat}{Flat}
\DeclareMathOperator{\Ch}{\cC{\mathsf h}}

\title{Remarks on the locality of generalized global symmetries}

\author{Owen Gwilliam}

\begin{document}

\maketitle

\begin{comment}
\epigraph{\dots  a mapping space $\Map(X; Q)$ is of the form $C(X; P)$ only when the target space $Q$ is $d$‐connected, 
and I would guess that an analogous distinction applies to field theories: 
just as the mapping space cannot be modelled by particles, 
but requires $m$‐dimensional ‘objects’, if $Q$ is only ($d$-$m$)‐connected, so we know that a field theory in general has non‐
local observables that can be seen only in topologically non‐trivial regions of space‐time.}{Graeme Segal from {\it Locality of Holomorphic Bundles},\\ {\it and Locality in Quantum Field Theory}}
\end{comment}

\epigraph{\dots we would have to allow not just ‘particles’ but also configurations of higher dimensional submanifolds \dots  a field theory in general has non‐local observables that can be seen only in topologically non‐trivial regions of space‐time.}{Graeme Segal from {\it Locality of Holomorphic Bundles},\\ {\it and Locality in Quantum Field Theory}}

The notion of generalized global symmetry is undergoing a period of rapid development, as it offers a novel perspective on rich phenomena in physics.
Mathematics also has a long tradition of generalizing the group concept --- the codification {\it par excellence} of symmetry --- and so a mathematical physicist will want to understand and formalize this new perspective from physics.

Here is the ``definition'' found in the paper \cite{GKSW} that initiated both this vocabulary of generalized global symmetry and also much of the activity in this direction:
\begin{quote}
a $q$-form symmetry in $d$ dimensions is implemented by an operator associated with a codimension $q+1$ closed manifold $M^{(d-q-1)}$,
\[
U_g(M^{(d-q-1)}),
\]
where $g \in G$ is an element of the symmetry group $G$. The fact that this is a symmetry means that the manifold $M^{(d-q-1)}$ can be deformed slightly without affecting correlation functions -- the answers depend only on the topology of $M^{(d-q-1)}$. Such operators can be multiplied
\[
U_g(M^{(d-q-1)}) U_{g'}(M^{(d-q-1)}) = U_{g''}(M^{(d-q-1)})
\]
with $g'' = g g' \in G$. As we will see below, for $q > 0$ the symmetry group $G$ must be Abelian.
\end{quote}
Later references offer quite similar descriptions~\cite{Snowmass, Bhardwaj,SchafNam,HarOog}.

A mathematician will likely find this description a bit incomplete,
as it is vague about some details,\footnote{For instance,  does ``deformed slightly'' mean isotopy or allow something more drastic? How does one combine $U_g(M)$ and $U_{g'}(M')$ for distinct submanifolds $M$ and $M'$? Is this somehow functorial in the submanifolds? Note as well that the definition involves the symmetry {\it acting} through something else, as if defining a group by its action on some other set. More intimidating still, the group is acting on a quantum field theory --- a notion that has notoriously evaded a satisfactory mathematical formalization.}
but will also feel it ought to fit inside algebraic topology,
if a little care is taken to fix the details.

And indeed it does: we'll see that a $q$-form symmetry operator is an element of the {\em compactly supported} cohomology group
\[
H^{q+1}_c(-,G).
\]
This perspective suggests, to an algebraic topologist, a refinement that contains a lot of interesting information:
take the {\it whole space}
\[
\Map_c(-,B^{q+1}G)
\]
where $B^{q+1}G$ denotes the Eilenberg-Maclane space~$K(G,q+1)$.
The compactly supported cohomology group~$H^{q+1}_c(-,G)$ is just the set of connected components $\pi_0\Map_c(-,B^{q+1}G)$, so the space knows a lot more information.

The goal of this paper is to take this suggestion and see how it brings together ideas from algebraic topology (notably factorization homology) and from physics (generalized global symmetries).

Now let's jump into reworking the definition of~\cite{GKSW} and hence to understanding our assertion above.

\subsection*{First pass at capturing $q$-form symmetry}

Fix a $d$-dimensional manifold~$X$ and an abelian group $A$.
(We switch from $G$ to $A$ to foreground the abelianness.)
Consider the functor
\[
\begin{array}{ccc}
\Open(X) & \to & \AbGrp\\
V & \mapsto & H_{d-(q+1)}(V,A)
\end{array}
\]
where 
\begin{itemize}
\item $\AbGrp$ denotes the category of abelian groups and group homomorphisms (also called group maps);
\item $\Open(X)$ denotes the partially ordered set of open subsets of the manifold $X$, viewed as a category,\footnote{We use this category $\Open(X)$ because it is simple and quite concrete, but one can adjust or generalize in many ways. 
If preferred, one can restrict to a special class of nice open subsets (such as open disks); 
one can instead work with all manifolds of dimension~$d$ and smooth embeddings, or with Riemannian $d$-manifolds (and isometric embeddings), or with all manifolds of any dimension, or something more general. 
This choice of category analogous to $\Open(X)$ is known as a {\em site}, and the optimal choice depends on the problem of interest.} and 
\item $H_{d-(q+1)}(V,A)$ denotes the $d-(q+1)$th homology of $V$ with coefficients in an abelian group~$A$ (e.g., constructed via singular chains).
\end{itemize}
Let's now see how this functor encodes the notion of a $q$-form symmetry valued in~$A$.

Pick a connected, closed codimension $q+1$ submanifold $M \subset X$ and then choose $V \supset M$ is a tubular neighborhood.
Observe that
\[
H_{d-(q+1)}(V,A) \cong H_{d-(q+1)}(M,A),
\]
and, moreover, that
\[
H_{d-(q+1)}(M,A) \cong A.
\]
In other words, an element of this group $H_{d-(q+1)}(V,A)$ can be seen as $M$ labeled by an element of~$A$.\footnote{\label{foot: or} We fess up here to an important condition: 
the second isomorphism requires that $M$ is orientable for the group~$A$. 
For instance, if $A = \ZZ$, then we need $M$ to be orientable in the usual sense, 
but if $A = \ZZ/2$, there is no condition on $M$. 
This orientability issue plays a role in later papers on generalized global symmetries, 
even if it is suppressed in the definition we quoted.}
Such an element matches quite explicitly with the  quotation from~\cite{GKSW}.\footnote{We note a subtlety here: 
not every homology class need be represented by a submanifold.
Instead, one allows more singular subspaces, usually built up from simplices mapped into~$X$. 
These can be viewed as more complicated configurations of defects for a theory, and we suspect that the authors of~\cite{GKSW} would accept $q$-form symmetries built in this manner.}

One nice feature of this functor $H_{d-(q+1)}(-,A)$ is precisely its functoriality: 
it tells us how to relate a $q$-form symmetry on one open set to a $q$-form symmetry on a larger open set.
We also learn how to combine elements from disjoint open sets.
Given $V, V'$ disjoint opens and both contained in $W$, there is a group map
\[
H_{d-(q+1)}(V,A) \oplus H_{d-(q+1)}(V',A) \cong H_{d-(q+1)}(V \cup V',A) \to H_{d-(q+1)}(W,A).
\]
Thus, if we shift from talking about submanifolds to working with open sets,
we obtain a more detailed and explicit description of $q$-form symmetry.

The reader might be wondering about this use of homology,
when we initially mentioned compactly-supported {\it co}\/homology.
The translation here is via Poincar\'e duality,
which provides a natural isomorphism
\[
H^{q+1}_c(V,A) \cong H_{d-(q+1)}(V,A)
\]
if $X$ is oriented. 
(Recall footnote~\ref{foot: or}.)

\subsection*{Enriching this model, like a topologist}

Experience suggests it might be fruitful to move beyond a fixed group $H_{d-(q+1)}(-,A) \cong H^{q+1}_c(-,A)$
and work with a richer structure.
We will simply introduce such an enhancement and then unwind how it fits with the story of generalized global symmetries as well as factorization algebras.

There is a nifty description of cohomology in a terms of mapping spaces, as follows.
Let $B^n A$ denote an Eilenberg-MacLane space $K(A,n)$ and recall that for a topological space $Z$, we have
\[
\pi_0 \Map(Z,B^n A) = H^n(Z,A),
\]
so that $B^n A$ is a space that represents the $n$th cohomology group with coefficients in~$A$.
Fix a basepoint $p$ in $B^n A$.\footnote{For instance, although we only care about the space $B^n A$ up to homotopy equivalence,
one can construct a representative among topological spaces that is a topological abelian group,
in which case the identity is a natural basepoint.}
Then one can ask about {\em compactly supported} continuous maps instead of all continuous maps:
\[
\Map_c(Z,B^n A) = \{ f: Z \to B^d A \,:\, f^{-1}(B^n A \setminus \{p\}) \text{ has compact closure}  \},
\]
i.e., the maps $f$ whose support is compact.
Compactly supported cohomology is hence
\[
H^n_c(Y,A) := \pi_0 \Map_c(Y,B^n A).
\]
In other words, there is a functor
\[
\begin{array}{ccc}
\Open(X) & \to & \Top\\
V & \mapsto & \Map_c(V,B^n A)
\end{array}
\]
where $\Top$ denotes the category of topological spaces and continuous maps,
and it provides an enhancement of $H^n_c(-,A)$ since we can recover that simpler functor by taking connected components~$\pi_0$.

In short, the functor $\Map_c(-,B^{q+1}A)$ offers a rich mathematical incarnation of $q$-form symmetries with coefficients in~$A$.\footnote{This approach also offers natural variations to handle subtleties, like orientability, such as by replacing maps into $B^n A$ by sections of a fiber bundle over $X$ with fiber~$B^n A$.}

We note that something special happens when $q = 0$ that lets us capture discrete symmetries of the traditional kind:
we can allow {\it non}\/abelian groups as well,
because there is a classifying space $BG = K(G,1)$ for a nonabelian discrete group~$G$.
(We discuss Lie groups and ``continuous'' generalized symmetries in Section~\ref{sec: smooth symmetries}.)
Here we model $0$-form symmetries valued in $G$ by the functor $\Map_c(-,BG)$.
To match with~\cite{GKSW},
consider $M$ a closed codimension~1 submanifold of $X$ whose normal bundle is trivializable
and let $V \supset M$ be a tubular neighborhood.
Then 
\[
\Map_c(V,BG) \simeq \Map_c(M \times \RR, BG) \simeq \Map(M,\Omega BG) \simeq \Map(M,G) \simeq G^{\pi_0(M)},
\]
so our functor returns $G$-valued functions on the connected components of~$M$.
In other words, $\Map_c(V,BG)$ consists of all possible 0-form operators supported on~$M$,
i.e., it is the set $\{U_g(M^(d-1)\,;\, g \in G\}$ for this fixed~$M$,
using the notation of~\cite{GKSW}.

\subsection*{Linearizing $q$-form symmetries}

Other interesting constructions arise by postcomposing with this functor.
For instance, the composite
\[
\Open(M) \xto{\Map_c(-,B^{q+1} A)}  \Top \xto{H_0(-,\ZZ)} \AbGrp
\]
assigns the group ring $\ZZ[A]$ to (a tubular neighborhood of) a connected, closed $(q+1)$-dimensional submanifold.\footnote{To see this, observe that discrete sets form a subcategory of topological spaces, 
and on this subcategory $H_0(-,\ZZ)$ agrees with the free abelian group functor. 
When $V$ is such a tubular neighborhood, then $H_{q+1}(V,A) \cong A$ is a group 
and so 
\[
H_0(H_{q+1}(V,A),\ZZ) \cong \ZZ[A]
\]
where the group structure on $A$ determines the multiplication on the free abelian group~$\ZZ[A]$.}
This functor provides a kind of ``noninvertible'' symmetry as most elements of the group ring are not invertible.

More information appears by taking the composite
\[
\Open(M) \xto{\Map_c(-,B^{q+1} A)}  \Top \xto{C_\bullet(-,\ZZ)} \AbGrp
\]
where $C_\bullet(-,\ZZ)$ denotes the singular chains,
which is a kind of ``dg group ring.''
Similarly, taking the composite
\[
\Open(M) \xto{\Map_c(-,B^{q+1} A)}  \Top_* \xto{\pi_d} \AbGrp
\]
with a higher homotopy group~$\pi_d$ captures more subtle information about the collection of $q$-form symmetries.

\subsection*{What happens in this paper}

A number of questions might occur to the reader at this point, such as
\begin{itemize}

\item What kind of mathematical object is this functor? What kind of properties does it have?

\item Can we encode higher group symmetries? That is, can we go beyond higher form symmetries?

\item Can we describe {\em how} generalized global symmetries act on a field theory using this framework?

\end{itemize}
Below we offer extended answers to the first two questions in Section~\ref{sec: core}.
This formulation of higher group symmetries has the virtue that it easily accommodates infinite groups (not just finite groups) and Lie groups, as a kind of unstable analog of differential cohomology theories.

In Section~\ref{sec: NAPD}, 
inspired by Segal~\cite{Segal},\footnote{We strongly encourage the reader to go back to this wonderful essay of Segal. 
This note is an homage to his vision.} 
we discuss the question of {\em locality}:
these higher group symmetries form factorization algebras (of a new kind),
using Grothendieck topologies designed to capture higher-dimensional defects.

The third question admits, at the moment, a less satisfying answer, largely because it is hard to capture quantum field theory with adequate mathematical precision in any framework.\footnote{It's fair to say that the functorial viewpoint, largely initiated by Atiyah and Segal, is highly successful for topological field theories 
but much less developed for examples of central physical interest, like Yang-Mills theories or gauge theories coupled to matter. 
There is thus a gap between the mathematical description of the higher symmetries abstractly (e.g., in \cite{FMT}) and how they act on such field theories.}
In Section~\ref{sec: action} we describe a point of view and suggest potential avenues of development.

Finally, in Section~\ref{sec: vistas} we sketch how one might incorporate {\em anomalies} and {\em continuous} generalized symmetries.

{\bf Caveat lector:} 
This paper does not examine specific examples --- the physics literature has many! --- and it does not include illustrations to develop intuition.
See \cite{Bhardwaj} for many lovely and illuminating pictures that accompany effective exposition of the intuition and ideas. 
Our goal is to offer a mathematical home for these examples.
For the mathematical reader, we emphasize that we do {\em not} use functorial field theories, 
and we do not offer an attempt to synthesize our approach  with mathematical approaches using functorial language, such as~\cite{FMT}.\footnote{One can start in this direction, following Lurie in Section 4.1 of \cite{LurieTFT} and Scheimbauer~\cite{Scheim}.}
We also acknowledge that this paper contains no proofs and many incomplete definitions\footnote{In light of the opening to this paper, I clearly cast stones in glassy houses.},
as it aims to outline a direction of possible mathematical development.

\subsection*{Acknowledgements}

This note was inspired by attendance at events of the Simons Collaboration on Global Categorical Symmetries.
which made me feel obligated to understand these ideas in my own terms.
Discussions there with David Ayala led to our joint work on generalizing factorization algebras and the nonabelian Poincar\'e duality theorem;
it's been a joy to work with David.
I thank the Simons Collaboration for fostering these efforts.

Many others deserve thanks for honing the ideas presented here and for feedback that much improved the exposition.
Conversations with Araminta Amabel around recasting \cite{FreedIntro} using factorization algebras were an important impetus in this direction.
John Huerta joined that dialogue and later gave helpful feedback on this note.
I also thank audiences at lectures in South Bend, Bonn,  Berlin, and Edinburgh for questions and suggestions that improved his ideas and explanations.
More recently, detailed feedback and encouragement from Ben Gripaios and Lukas M\"uller were incredibly helpful, and I am quite grateful for it!

As usual for me, retrospectively I see that Kevin Costello had offered many key insights about these topics, 
and so Kevin deserves much credit for what's good here and I deserves any blame for poorly interpreting our discussions.

The National Science Foundation supported the author through DMS Grant 2042052. 
The Max Planck Institute of Mathematics offered a wonderful place to speak and write about these ideas, during a lovely and productive visit there, and I acknowledge its support with gratitude.

\tableofcontents

\section{Generalized symmetries as prefactorization algebras}
\label{sec: core}

This section is devoted to answering the first question about what kind of mathematical object we are seeing,
which then gives a straightforward way to encode higher group symmetries.
Before jumping into definitions, we offer a few orienting remarks.

Algebraic topologists have long studied functors like $\Map_c(-,Y)$;
as we'll discuss below, they helped spawn the theory of operads to capture rich structure beneath the surface of higher homotopy groups.
More recently,
topologists explored such functors in the setting of {\em factorization homology},
a kind of homology theory for manifolds that allows nonabelian coefficients,\footnote{Factorization homology is a pairing between $n$-dimensional manifolds (framed or oriented or equipped with some tangential structure) and $E_n$-algebras (if framed, otherwise modified to work with the tangential structure). It satisfies analogs of the Eilenberg-Steenrod axioms, notably a $\otimes$-version of excision. See \cite{AF} and \cite{LurieHA}, although Lurie calls it ``topological chiral homology.''}
which appear naturally in studying fully extended topological field theories:
for a beautiful introduction, see Section 4.1 of Lurie's exposition on the cobordism hypothesis \cite{LurieTFT}.  
In fact, Scheimbauer's thesis \cite{Scheim} proves that factorization homology gives an effective construction of a large class of fully extended theories.

In a moment we will introduce a broader context, namely the notion of a prefactorization algebra,
which arose as a very minimalistic encoding of the observables of a QFT (see \cite{CGEMP} for a survey of this topic).
We will show that the higher form symmetries provide examples of prefactorization algebras.

\subsection{Prefactorization algebras and higher form symmetries}

A key property of our examples is a kind of ``factorization'' property,
which we now describe.
Fix a $d$-dimensional manifold $X$ and a pointed, connected space $Y$ (we've usually taken an Eilenberg-Maclane space $Y= B^nA$).
Consider the functor 
\[
\Map_c(-,Y): \Open(X) \to \Top.
\]
If $U$ and $U'$ are disjoint open sets,
then
\[
\Map_c(U \sqcup U',Y) \cong \Map_c(U,Y) \times \Map_c(U',Y)
\]
because a map $f$ with support in $U \sqcup U'$ can be understood in terms of what it does on $U$ and $U'$ separately.
(Outside those opens, $f$ sends everything to the basepoint of~$Y$.)
This factorization lets us define a kind of ``multiplication'':
consider the composite map
\begin{equation}
\label{eqn: mult}
\Map_c(U,Y) \times \Map_c(U',Y) \cong \Map_c(U \sqcup U',Y) \to \Map_c(V,Y)
\end{equation}
for any $V$ containing both $U$ and~$U'$.
This composite map lets us combine (or ``multiply'') a map on $U$ with a map on $U'$ to get a map on $V$.
In this sense, it resembles the operator product in QFT.
(The picture in the next definition will display that analogy clearly.)

To capture this kind of behavior in structural languge, 
we now give the definition of a prefactorization algebra.
We take the target category to be topological spaces,
but it is possible (and straightforward) to replace it by any symmetric monoidal category or $\infty$-category.
After the definition, we will relate to $E_d$ algebras, 
also known as {\it little $d$-disks algebras}.

\begin{dfn}
Let $X$ be a topological space.  
A {\em prefactorization algebra} $\cF$ on $X$ with values in topological spaces $\Top$ is the following data:
\begin{itemize}
\item a topological space $\cF(U)$ for each open set $U \subset X$; 
\item  a continuous map $m_V^U: \cF(U) \rightarrow \cF(V)$ for each inclusion $U \subset V$ of open sets; and
\item a continuous map $m_V^{U_1,\ldots,U_n} : \cF(U_1) \times \cdots \times \cF(U_n) \rightarrow \cF(V)$ for every finite collection of open sets where each $U_i \subset V$ and where the $U_i$ are pairwise disjoint. 
The following picture represents the situation.
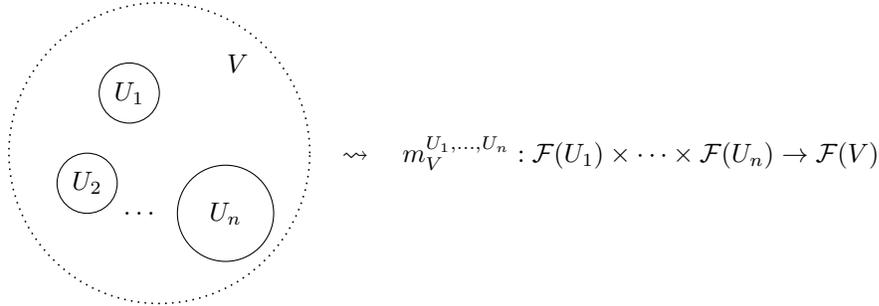
\begin{center}
 \begin{tikzpicture}[scale=0.8]
 \draw[dotted,semithick] (0,0) circle (2.5);
 \draw (-0.5,1) circle(0.5) node {$U_1$};
 \draw (-1.2,-0.5) circle (0.5) node {$U_2$};
 \draw (-0.3, -1) node {\dots};
 \draw (1.1,-1) circle (0.8) node {$U_n$};
 \draw (1.3, 1.5) node {$V$};
\node at (7.5,0){$\rightsquigarrow \quad m_V^{U_1,\ldots,U_n}: \cF(U_1)\times\cdots\times\cF(U_n)\to\cF(V)$};
 \end{tikzpicture}
\end{center}
\end{itemize}
The maps must be compatible in the obvious way, so that if $U_{i,1}\sqcup\cdots\sqcup U_{i,n_i}\subseteq V_i$ and $V_1\sqcup\cdots\sqcup V_k\subseteq W$, the following diagram commutes:
\begin{center}
\begin{tikzcd}[column sep=small]
{\prod}^{k}_{i=1}{\prod}^{n_i}_{j=1}\cF(U_j) \arrow{dr} \arrow{rr} &   &{\prod}^k_{i=1}\cF(V_i) \arrow{dl}\\
&\cF(W)  &
\end{tikzcd}.
\end{center}
\end{dfn}

For an explicit example of the associativity, consider the following picture.
\begin{center}
\begin{tikzpicture}[scale=0.9]
% big circle:
\draw[semithick,dotted] (0,0) circle (2);
\draw (1.2, 1.2) node {$W$};
 % circle V_1:
\draw [style= dashed,rotate=45] (-0.1,0.85) ellipse (1.5 and 0.9);
\draw (-0.9, 1.1) node {$V_1$};
 % circle V_2:
\draw [style= dashed] (0.9,-0.5) ellipse (.7 and 1);
\draw (0.9, 0) node {$V_2$};
 % small circles:
\draw (-1, 0.2) circle (0.6) node {$U_{1,1}$};
\draw (-0.1, 1.1) circle (0.5) node {$U_{1,2}$};
\draw (0.9,-0.9) circle (0.4) node {$U_{2,1}$};
%\draw (2.9, -2.1) circle (1.1) node {$U_{2,2}$};
\node at (6.5,0){$\rightsquigarrow
\begin{tikzcd}[column sep=small]
\cF(U_{1,1}) \times \cF(U_{1,2}) \times \cF(U_{2,1}) \arrow{dr} \arrow{d}  & \\ 
\cF(V_1) \times \cF(V_2) \arrow{r} &\cF(W) 
\end{tikzcd}
$};
\end{tikzpicture}\\
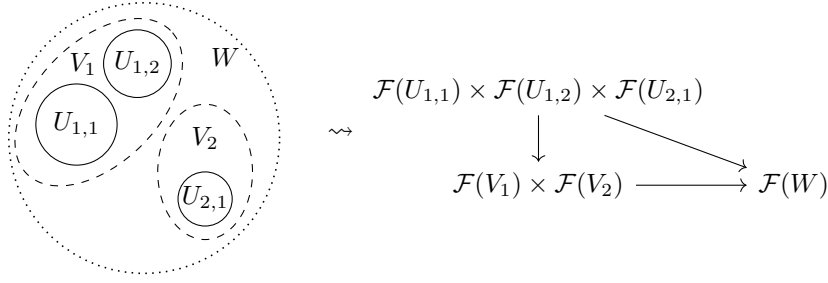
The case of $k=n_1=2$, $n_2 = 1$
\end{center}

This definition aims to formalize the fundamental properties that the operators of a QFT satisfy in terms of their support.

We have seen already that the functor $\cF = \Map_c(-,Y)$ provides an example,
by the discussion about the ``multiplication'' map~\eqref{eqn: mult}.
In particular, we now see how $q$-form symmetries fit into this framework.

\begin{dfn}
Let $X$ be a $d$-manifold, $A$ an abelian group, and $d \geq 1$.
The {\em $q$-form symmetry algebra} on $X$ valued in $A$ is the prefactorization algebra~$\Map_c(-,B^{q+1}A)$.
\end{dfn}

When $q = 0$, we can work with a discrete nonabelian group~$G$ to capture traditional discrete symmetries.

\begin{rmk}
In Section~\ref{sec: higher group} we explain how to interpret the prefactorization algebra $\Map_c(-,Y)$ for arbitrary pointed space~$Y$, 
rather than just Eilenberg-Maclane spaces~$Y = B^{q+1}A$.
\end{rmk}

Algebraic topologists have studied the general case $\Map_c(-,Y)$ for a long time,
but using different terminology.
Consider, in particular, the case where the source manifold is $X = \RR^d$.
Then
\[
\Map_c(\RR^d, Y) \cong \Omega^d Y,
\]
the $d$-fold based loop space of~$Y$,
whose connected components are
\[
\pi_0(\Omega^d Y) = \pi_d(Y)
\]
and have an interesting abelian group structure.
The space $\Omega^d Y$ itself has an enhancement of this abelian group structure,
and to describe the full algebraic structure on $\Omega^d Y$,
topologists developed the framework of $E_d$ algebras,
where $E_d$ is the operad of little $d$-disks.
In the case $d=1$, where we are discussing composition of loops,
this operad $E_1$ is also known as $A_\infty$,
since the based loops $\Omega Y$ are associative up to coherent homotopy but not strictly associative.

The $E_d$ multiplication maps are encoded in our functor $\Map_c(-,Y)$ if we consider its behavior on the subcategory $\Disk(\RR^d) \subset \Open(\RR^d)$ consisting of open $d$-dimensional disks sitting inside $\RR^d$.
Then the multiplication maps of our prefactorization algebra encode how to take products of elements in~$\Omega^d Y$.
For this reason, prefactorization algebras offer a generalization of $E_d$ algebras,
as they allow for examples that are not locally constant along~$X$.
See \cite{CGEMP} for an introduction to their role in QFT.

\begin{rmk}
Such $E_d$ algebras appear naturally in topological field theories too:
given a $d$-dimensional TFT $Z$ with values in a category $\cC^\otimes$,
the value $Z(S^{d-1})$ is an $E_d$ algebra in~$\cC^\otimes$.
\end{rmk}

\subsection{Higher group symmetries}
\label{sec: higher group}

\begin{comment}
In this section we first move from higher form symmetries to higher group symmetries,
which amounts to replacing an Eilenberg-Maclane space $K(A,n)$ as the target with an arbitrary pointed space~$Y$.
This shift amounts to allowing an $\infty$-groupoid of symmetries that has support on submanifolds of varying dimension, 
rather than just a group labeling submanifolds of a fixed dimension.

\end{comment}

We have seen that the functor $\Map_c(-,Y)$ is a prefactorization algebra for any pointed space $Y$,
and we might wonder whether it has any meaning akin to the higher form symmetry prefactorization algebras for Eilenberg-Maclane spaces.
A topologist might say that any pointed space $Y$ can be seen as a kind of $\infty$-groupoid\footnote{In fact, if $Y$ is connected, then there is a homotopy equivalence $Y \simeq B\Omega Y$, and so $\Map_c(-,B\Omega Y)$ is a version of generalized symmetries with the ``group'' $\Omega Y$.} and so would be optimistic that $\Map_c(-,Y)$ must encode some kind of generalized symmetry.
We now justify the topologist's optimism, 
matching it with physical terminology.

In the physics literature a {\em higher group symmetry} arises when higher form symmetries of different degrees mix.
For instance, one might have a 1-form symmetry whose action affects a 3-form symmetry.
In practice, physicists often talk about extending a $q$-form symmetry by a $q'$-form symmetry and invoke computations about group extensions.
(See, e.g., Section 2.3 of \cite{Snowmass} or Section 5 of~\cite{Bhardwaj}.)

Given that a higher form symmetry is encoded by a prefactorization algebra,
we might ask how to ``mix'' degrees or ``extend'' using topological tools.
The key is to broaden one's notion of symmetry beyond groups.
Here we sketch two approaches to broadening one's vision.

First, every space can be viewed as built in layers from Eilenberg-Maclane spaces.
Suppose that we start with a pointed space $Y$ that is path-connected.
Then there is a {\em Postnikov tower}
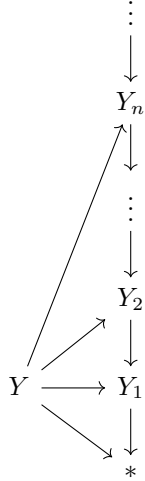
\[
\begin{tikzcd}
& \vdots \ar[d]\\
& Y_n \ar[d] \\
& \vdots \ar[d] \\
& Y_2 \ar[d]\\
Y \ar[dr] \ar[r] \ar[ur] \ar[uuur]  & Y_1 \ar[d] \\
& *
\end{tikzcd}
\]
where each stage $Y_n$ only has nonzero homotopy groups in dimensions~$\leq n$, where any two consecutive layers form a fibration
\[
\begin{tikzcd}
K(\pi_n(Y),n) \ar[r] & Y_n \ar[d, "p_n"] \\
& Y_{n-1}
\end{tikzcd},
\]
and 
\[
Y \simeq \lim_{\longleftarrow} Y_n,
\]
i.e., $Y$ is weakly homotopy equivalent to the limit of this tower.

This tower describes $Y$, as a homotopy type, in essentially algebraic terms:
\begin{itemize}
\item each Eilenberg-Maclane space can be viewed as a kind of group,
\item each layer $Y_n$ can be viewed as an {\em extension} of the ``group'' $Y_{n-1}$ by the group~$K(\pi_n(Y),n)$, and
\item each fibration $p_n$ is classified by a cohomology class $H^{n+1}(Y_{n-1}, \pi_n(Y))$, known as a Postnikov invariant.
\end{itemize}
In other words, $Y$ is assembled out of a sequence of groups and group extensions, in some generalized sense.

This ``layer cake'' philosophy \cite{BaezShul} lets us view $\Map_c(-,Y)$ as assembled from higher form symmetry algebras $\Map_c(-,K(\pi_n(Y),n))$ using the Postnikov invariants.
Concretely, we know
\begin{align*}
\Map_c(-,Y) 
&\simeq \Map_c(-,\lim_{\longleftarrow} Y_n) \\
&\simeq \lim_{\longleftarrow} \Map_c(-,Y_n)
\end{align*}
so we can analyze the prefactorization algebra $\Map_c(-,Y)$ as a sequence of extensions of higher form symmetries.

This perspective, via Postnikov towers, suggests the following definition.

\begin{dfn}
Let $X$ be a $d$-manifold and let $Y$ be a pointed, connected space. 
The {\em higher group symmetry algebra} on $X$ and valued in~$Y$ is the prefactorization algebra~$\Map_c(-,Y)$.
\end{dfn}

There is a second perspective on why a space $Y$ encodes a kind of symmetry:
a topological space can be seen as presenting an $\infty$-category 
whose objects are its points, 
whose 1-morphisms are paths between points, 
whose 2-morphisms are homotopies between paths, and so on.
Recall that a category is a ``horizontal'' generalization of a monoid,
and a group is a monoid with inverses.
Hence a topological space determines a special kind of $\infty$-category where everything has inverses,
known as an $\infty$-groupoid.

From this point of view, $\Map_c(-,Y)$ parametrizes ways to produce elements of this $\infty$-groupoid but labeled by compact regions of spacetime.
Alternatively, in the style of \cite{GKSW}, 
we are allowing an $\infty$-groupoid of symmetries that has support on submanifolds of varying dimension, 
rather than just a group labeling submanifolds of a fixed dimension.

\begin{rmk}
One can move beyond an $\infty$-groupoid of symmetries to higher categories, using this factorization perspective.
The original work in this direction is the {\em blob homology} of Morrison and Walker \cite{MorWal},
and there is the $\beta$-factorization homology of Ayala-Francis-Rozenblyum that generalizes factorization homology to allow $(\infty,n)$-categories as coefficients, not just $E_n$ algebras~\cite{AFR}.
\end{rmk}

\section{A new notion of factorization algebra and a novel version of nonabelian Poincar\'e duality}
\label{sec: NAPD}

Field theories are local-to-global in nature.
For classical field theories, this feature is manifest:
the equations of motion are PDE, so they impose a very local condition and solutions can be patched together on covers.
In mathematical language, we say that solutions to the equations of motion form a sheaf on spacetime.\footnote{We mean here solutions without some global regularity condition, such as smooth or distributional distributions. Such regularity conditions, like $L^2$, are often not local in nature.}

One might hope, then, that generalized symmetries also satisfy some local-to-global property.

Note, for example, that the functor 
\[
\Map(-,Y): \Open(X)^\op \to \Top
\]
is a sheaf, 
since you can reconstruct continuous maps on a big open by taking continuous functions on a cover that agree on overlaps.
Our functor 
\[
\Map_c(-,Y): \Open(X) \to \Top
\]
is very similar but covariant,
so we might guess that it is a {\em co}\/sheaf.
(A cosheaf is a covariant functor out of $\Open(X)$ that satisfies the ``dual'' of the sheaf condition.)\footnote{Let $\cF$ be a sheaf. Given a cover $\{U_i\}$ of $V$, then $\cF(V)$ equalizes the diagram $\prod_i \cF(U_i) \rightrightarrows \prod_{j,k} \cF(U_j \cap U_k)$. For a cosheaf $\cA$, $\cA(V)$ {\em co}\/equalizes the diagram $\sqcup_{j,k} \cA(U_j \cap U_k) \rightrightarrows \sqcup_i \cA(U_i)$. For discussions of cosheaves, see \cite{Bredon, Curry, CG1}.}

\begin{comment}
These examples do not, typically, form {\em factorization} algebras in the sense of~\cite{CG1,CG2} and hence fall outside that formalism, 
which was developed to capture perturbative QFT.
At the end of this section, we indicate a natural generalization that does seem to capture such nonperturbative phenomena.
\end{comment}

Alas, our examples are not cosheaves for the usual topology on a manifold $X$, because
\[
\Map_c(U \sqcup U',Y) \cong \Map_c(U,Y) \times \Map_c(U',Y)
\]
for disjoint opens $U, U'$, whereas a cosheaf $F$ would satisfy
\[
F(U \sqcup U') \cong F(U) \sqcup F(U').
\]
Thankfully, there is a remarkably simple family of Grothendieck topologies for which our functors {\em are} cosheaves.

\begin{dfn}
\label{dfn: ksupp}
For each natural number $0 \leq k \leq d$, let $\cT_k[X]$ denote the Grothendieck topology on $\Open(X)$ generated by covers $\cU = \{U_i\}_{i \in I}$ with the property that for any inclusion of a finite cell complex $K$ of dimension $\leq k$,\footnote{We mean here that $K$ is built from a finite collection of cells $\{e^r_\alpha\}$ where $\dim(e^r_\alpha) = r \leq k$ holds for all~$\alpha$.} 
there is some $U_i$ containing~$K$.\footnote{This definition works well for topological manifolds, and there are suitable variants for PL and smooth manifolds. Theorem~\ref{NAPD} is proved for PL manifolds using the PL version, as it is technically convenient to use triangulations.}
We call $\cT_k[X]$ the $k$-{\em dimensionally supportive topology}, as it contains all $k$-dimensional defects.
\end{dfn}

For $k = 0$, this 0d-supportive topology $\cT_0[X]$ says that any finite set of points in $X$ is contained in some open set $U_i$ from the cover.
It is known as the Weiss topology, already developed in the literature of manifold calculus and factorization algebras.
For $k = 1$, this 1d-supportive topology says that the every embedded {\em graph} is contained in some open set of the cover.
Note that these topologies demand covers of increasingly large size.

These topologies encode a clear physical idea: 
to reconstruct $k$-dimensional defects, a cover must contain a tubular neighborhood around any defect of dimension at least~$k$.
When $k=0$, this topology says that a cover knows about all the 0-dimensional defects,
and hence it captures the old saw that if one knows all the $n$-point functions, then one can reconstruct the QFT.

The following theorem justifies this physical idea in the context of higher group symmetries, 
and we view it as tentative evidence that these topologies are relevant to field theory in general.\footnote{Recall the epigraph of Segal, 
whose \cite{Segal} offers an elegant overview of the ideas informing this work. 
That paper concludes: ``a mapping space $\Map(X; Q)$ is of the form $C(X; P)$ only when the target space $Q$ is $d$‐connected, 
and I would guess that an analogous distinction applies to field theories: 
just as the mapping space cannot be modelled by particles, 
but requires $m$‐dimensional ‘objects’, if $Q$ is only ($d$-$m$)‐connected, so we know that a field theory in general has non‐local observables that can be seen only in topologically non‐trivial regions of space‐time.''}

\begin{thm}[\cite{AG}]
\label{NAPD}
Let $X$ be a $d$-manifold, and $Y$ a pointed space.
If $Y$ is ($d$-$k$)-connective (i.e., $\pi_n(Y) = 0$ for $n < d-k$),
then the functor 
\[
\Map_c(-,Y): \Open(X) \to \Spaces
\]
is a cosheaf for the $\cT_k[X]$ topology.
\end{thm} 

\begin{rmk}
When $k=0$, this theorem is a consequence of the {\em nonabelian Poincar\'e duality} of Salvatore, Lurie, and others, and is essentially equivalent.\footnote{The history here is a bit complicated, as mapping spaces are a central subject in topology.
There are many comparison results between labeled configuration spaces, factorization algebras, factorization homology, topological chiral homology, and the scanning map \cite{McDuffSegal, AF,LurieHA,Miller, KSW}.}
To see the connection with usual Poincar\'e duality, consider the special case where $X$ is a $d$-manifold and $Y = B^d(A) = K(A,d)$,
with $A$ an abelian group.
In this case $Y$ is $d$-connective so we use the $\cT_0[X]$ topology,
which is the Weiss topology.
Note that $\pi_0(\Map_c(X,K(A,d)))$ is the compactly-supported cohomology $H^d_c(X,A)$.
The theorem above tells us we can reconstruct this information by knowing the functor $\Map_c(-,K(A,d))$ just on open disks in $X$,
i.e., using the Weiss cover of finite disjoint unions of disks.
But
\[
\Map_c(\RR^d,K(A,d)) \simeq A,
\]
so computing the colimit over this Weiss cover of disks recovers $H_0(X,A)$,
if $X$ is $A$-oriented.
Thus
\[
H^d_c(X,A) \cong H_0(X,A)
\]
as claimed by ordinary Poincar\'e duality.
\end{rmk}

This result also indicates which topology captures $q$-form symmetries on a $d$-manifold.

\begin{crl}
Let $X$ be a $d$-manifold, and let $A$ be an abelian group.
For $q +1 \leq d$, the $q$-form symmetry algebra $\Map_c(-,B^{q+1}A)$ is a cosheaf for the $\cT_{d-q-1}[X]$-topology.
\end{crl}

In other words, it is a $(d-q-1)$d{\em-supportive factorization algebra}.

\section{How generalized global symmetries act on theories}
\label{sec: action}

Although these factorization algebras capture generalized symmetries effectively in the {\em abstract}, 
much as a group captures the abstract notion of ordinary symmetry,
we would like to see how these factorization algebras {\em act} as symmetries of field theories,
to realize the initial vision of~\cite{GKSW}.
It would be easiest if the observables (or operators) of a QFT also formed a factorization algebra,
so that both symmetries and operators lived in the same setting.
Formulating QFT in factorization terms is close in spirit to the algebraic QFT of Haag and Kastler~\cite{HK, Haag},\footnote{The survey article~\cite{CGEMP} explicates factorization algebras as a rather minimal set of axioms of QFT and compares them to other structural approaches, including AQFT and functorial QFT.} 
and it works quite effectively for perturbative QFT~\cite{CG2}.

In this section we will seek here to cast generalized global symmetries acting on a QFT in a factorization framework.
In fact, we offer two approaches, 
inspired by positive results in the setting of TFT or perturbative QFT.

First, we will describe here a proposal from \cite{CGEMP} for a generalized Noether theorem as a map from a factorization algebra of higher group symmetries to a factorization algebra of observables for a nonperturbative field theory.
It is modeled on a factorization Noether theorem for pertubative field theories, 
mapping a $L_\infty$ algebra of symmetries to a factorization algebra of observables for a perturbative field theory, 
proven in~\cite{CG2}.
The perturbative version uses the 0d-supportive topology (i.e., the Weiss topology),
while the nonperturbative version would use the $k$d-supportive topology relevant for the higher group symmetry algebra and the algebra of observables.

Second, we indicate how the SymTFT, involving open-faced sandwiches~\cite{SchafNam,FMT}, can be articulated using factorization algebras.
This point of view fits nicely with describing a bulk-boundary field theory using a factorization algebra on a manifold with boundary,
and with the emerging theory of centralizers (or commutants) for factorization algebras~\cite{LurieHA, Francis}.

\subsection{As currents}

We begin by sketching why classical observables for a field theory form a prefactorization algebra so that we can formulate a map {\it \`a la} Noether.

Any Lagrangian field theory comes with a natural sheaf on the spacetime manifold~$X$:
typically, the fields are sections of a fiber bundle $F$ over $X$ so let
\[
\cF: \Open(X)^{\op} \to {\dStk}
\]
assign to each open $U \subset X$,
the sections of $F$ over~$U$.
Note that we use a target $\infty$-category ${\dStk}$ of ``derived stacks'' to accommodate theories with sophisticated or exotic fields (e.g., a gauge theory might be viewed as taking values in stacks, if one takes the stacky quotient of connections by gauge transformations).
We will be vague about the details of this $\infty$-category because a judicious choice might depend on the theory.\footnote{See, however, \cite{Steffens23, AlfonsiYoung, Steffens24} for work towards a good setting to do Lagrangian field theory with derived geometry.}
For a classical field theory, there is a subsheaf $\Sol$ of $\cF$ consisting of solutions to the equations of motion.

Every derived stack $S$ has a dg commutative algebra $\cO(S)$ of functions,
and taking functions defines a functor $\cO: {\dStk}^{\op} \to \CAlg_\CC$.
Observe that there is a natural map
\begin{equation}
\label{eqn: mult obs}
\cO(S) \otimes \cO(T) \to \cO(S \times T)
\end{equation}
for any derived stacks $S$ and $T$ by multiplying the functions pulled back along the projection maps.

The observables of a classical field theory on spacetime $X$ should be $\cO(\cF(X))$, 
the functions on the space of solutions.

\begin{dfn}
The {\em observables} are the composite functor
\[
\Obs: \Open(X) \xto{\Sol} \dStk^{\op} \xto{\cO} \CAlg_\CC
\]
that assigns ``observables with support in $U$'' to an open set~$U \subset X$.
\end{dfn}
  
For any disjoint opens $U,U'$ inside $V$,
there is a natural map
\[
\Obs(U) \otimes \Obs(U') \to \Obs(V)
\] 
due to the multiplication map~\eqref{eqn: mult obs} and the restriction map
\[
\Sol(V) \to \Sol(U \sqcup U') \cong \Sol(U) \times \Sol(U').
\]
This feature guarantees that $\Obs$ is a prefactorization algebra on~$X$.
We will view it as taking values in $\Ch_\CC$,
the $\infty$-category of dg complex vector spaces (up to quasi-isomorphism).\footnote{Is $\Obs$ a cosheaf for some kind of $k$-dimensionally supportive topology?}

We have already met another such prefactorization algebra  as the ``dg group algebra'' of a higher group symmetry:
\[
\Open(X) \xto{\Map_c(-,Y)} \Spaces \xto{C_*(-,\CC)} \Ch_\CC
\]
where $C_*(-,\CC)$ means singular chains valued in the complex numbers.

We can now formulate what it means for a theory to have a higher group symmetry.

\begin{dfn}
Let $X$ be a manifold, let $\Sol$ be a sheaf of solutions for a classical field theory, and let $Y$ be a pointed space.
A {\em current map} for higher $Y$-symmetries is a map
\[
J: C_*(\Map_c(-,Y),\CC) \to \Obs
\]
of prefactorization algebras on~$X$.
\end{dfn}

Such a map would offer a ``higher group'' analog of the Noether currents for infinitesimal symmetries.

Let's unwind what a current map would mean if $Y = B^{q+1}A$ and see how it compares with the notion from~\cite{GKSW}.
Let $M \subset X$ be a codimension $q+1$ submanifold that is closed and connected.
Pick a tubular neighborhood $V \supset M$.
Then we have a cochain map
\[
J(V): C_*(\Map_c(V,Y),\CC) \to \Obs(V)
\]
and, at the level of cohomology in degree~0, a linear map
\[
\CC[H_{d-(q+1)}(M,A)] \cong \CC[A] \to H^0 \Obs(V)
\]
from the group algebra of $A$ to the ``usual'' observables (in the sense that they sit in degree~0).\footnote{Recall that 
\[
\pi_0 \Map_c(V,B^{q+1}A) = H^{q+1}_c(V,A) \cong H_{d-(q+1)}(V,A) \cong H_{d-(q+1)}(M,A)
\]
using Poincar\'e duality in the second step and a deformation retraction of $V$ onto $M$ in the last step.}
In other words, each element $a \in A$ determines some element $U_a(M^{(d-q-1)})$ among the observables.

We have offered a formulation using {\em classical} observables, 
but if one could give a precise characterization of quantum observables,
then they should also form a prefactorization algebra.
For instance, in the Batalin-Vilkovisky quantization formalism,
one asks to deform $\Obs$ on each open set by deforming the differential.
This approach works well for perturbative QFT.

For more discussion of this circle of ideas, 
including 't Hooft anomalies and some examples, see the last section of the survey~\cite{CGEMP}.

\subsection{As a bulk-boundary system}

There is another perspective on how generalized symmetries appear in field theories:
a bulk-boundary system where a bulk field theory encodes the generalized symmetries  and couples to a theory on the boundary.
(People often talk about open-faced sandwiches or quiches in contemporary literature.)
Much of the analysis of generalized symmetries in QFT can be organized in this language (see, e.g., \cite{FMT}).\footnote{Note that there may be many different bulk theories that can couple to a given boundary theory.}

\begin{comment}
A similar situation arises in the setting of pure algebra. 
At the simplest level, consider a vector space $V$, parallel to the role of the boundary theory.
A symmetry of $V$ is an invertible endomorphism,
and a group $G$ acts by symmetries on $V$ means there is a group map $G \to \GL(V)$.\footnote{If the vector space has more structure, then ask for structure-preserving automorphisms. 
For example, if $V$ is a Hilbert space, replace $\GL(V)$ by the unitary group~${\rm U}(V)$.}
Thus $\GL(V)$ is the universal symmetry group of $V$ in the sense that any action on $V$ factors through it.
``Noninvertible'' symmetries can be understood as an algebra $A$ acting on $V$,
which is encoded by an algebra map $A \to \End(V)$,
so that $\End(V)$ is the universal symmetry algebra.
\end{comment}

The language of (pre)factorization algebras offers an easy way to encode the open-faced sandwich idea. 
Given a prefactorization algebra $\cA$ on a spacetime manifold $X$ that encodes a field theory,
one can ask for a prefactorization algebra $\cB$ on a thickening $X \times [0,t)$ such that 
\[
\cB(U \times [0,r)) \cong \cA(U)
\]
for any open $U \subset X$.
This ``bulk'' prefactorization algebra $\cB$ captures the open-faced sandwich story:
for $0 < r < s < t$, 
we have a structure map
\[
\cB(U \times (s,t)) \otimes \cB(U \times[0,r)) \to \cB(U \times [0,t))
\] 
and hence an action 
\[
\cB(U \times (s,t)) \otimes \cA(U) \to \cA(U)
\]
on the observables $\cA(U)$ by bulk operators in~$B(U \times (s,t))$.

Such bulk-boundary systems in a perturbative setting have been exhibited and analyzed using factorization algebras~\cite{GRW,RabThesis,ButsonYoo}.

One might posit that there is a {\em universal} bulk theory,
encoding the largest possible higher symmetry group,
and the SymTFT of a field theory corresponds, roughly, to this universal bulk theory.
It is natural then to ask how a universal bulk theory might be encoded as a prefactorization algebra.\footnote{A similar situation arises in the setting of pure algebra. 
At the simplest level, consider a vector space $V$, playing the role of the boundary theory (i.e., the factorization algebra $\cA$ with $X$ just a point).
A symmetry of $V$ is an invertible endomorphism,
and a group $G$ acts by symmetries on $V$ means there is a group map $G \to \GL(V)$.
Thus $\GL(V)$ is the universal symmetry group of $V$ in the sense that any action on $V$ factors through it.
``Noninvertible'' symmetries can be understood as an algebra $R$ acting on $V$,
which is encoded by an algebra map $R \to \End(V)$,
so that $\End(V)$ is the universal symmetry algebra.} 
In other words, we ask:
given a prefactorization algebra $\cA$ on $X$,
is there a bulk factorization algebra $\cB_{\rm univ}$ through which any other bulk factorization algebra $\cB$ acts on~$\cA$?\footnote{In the perturbative setting, see \cite{ButsonYoo} for a framework and several illuminating examples.}

Although there is not yet an answer at this level of generality,
it {\em is} answered when $\cA$ corresponds to a topological field theory.
More precisely, suppose $X = \RR^n$ and let $\cA$ be locally constant.
In this situation, $\cA$ amounts to an algebra over the little $n$-disks operad~$E_n$.
Here the answer is known to be yes:
for every $E_n$ algebra $A$, 
there is a universal $E_{n+1}$ algebra $Z(A)$ known as its {\em derived center}\footnote{A generalization of the Hochschild cohomology of an associative or $E_1$ algebra.},
and any $E_{n+1}$ algebra $B$ acting on an $E_n$ algebra $A$ must act through its derived center~$Z(A)$.
Deligne conjectured the existence of this derived center, with its $E_{n+1}$ structure;
it is now a theorem proved in many different ways.
Kontsevich conjectured the second statement, now also a theorem,
and he articulated it using Swiss cheese algebras,
which can be seen as factorization algebras on the half-space 
\[
\HH^{n+1} = \RR^n \times [0,\infty)
\]
where an $E_n$ algebra $A$ lives on the boundary and an $E_{n+1}$ algebra $B$ acting on $A$ lives in the bulk.
See \cite{Thomas, DTT} for proofs of Kontsevich's conjecture and see \cite{Horel} for a discussion in the framework of factorization algebras.

Inspired by these results, we make the following suggestion. 

\begin{expn}
Every prefactorization algebra encoding the observables of a QFT (possibly nonperturbative) has a universal bulk prefactorization algebra.
\end{expn}

We posit that the bulk theory encoded by this universal bulk prefactorization algebra should capture the SymTFT of the theory encoded by the ``boundary'' prefactorization algebra.

\section{Next steps}
\label{sec: vistas}

Here we offer tentative ideas about 
how to move beyond discrete groups in generalized symmetries (and hence pure algebraic topology) 
and about how to encode anomalous generalized symmetries.

\subsection{A suggestion about smooth higher symmetries}
\label{sec: smooth symmetries}

The framework we have introduced is easily modified to move beyond purely topological issues.
In particular we would like to talk about generalized symmetries involving Lie groups that depend on their manifold structure, 
not just on their underlying group structure.
For example, there is a difference between $U(1)$ with the topology as a circle versus $U(1)$ equipped with the discrete topology.
(Sometimes physicists use the phrase {\em continuous} generalized symmetry to refer, in essence, to this distinction.)
In the setting of algebraic topology, 
such a ``smooth refinement'' of ordinary cohomology leads to differential cohomology, 
which interfaces naturally with differential geometry, 
and it has played a big role in analyzing nonperturbative aspects of quantum field theories and string theories.\footnote{See \cite{WittenM,MooreWitten,FreedDCQ,FSS,HopkinsSinger} as jumping off points into the vast literature on differential cohomology theories and physics.}

Our approach uses compactly supported mapping spaces,
so naively one might ask to use spaces of compactly supported {\em smooth} maps into some pointed target space.
The challenge here is that classifying spaces and Eilenberg-Maclane spaces are rarely modeled by finite-dimensional smooth manifolds, 
so one must work a bit to find a suitable mathematical context.

We now sketch a potential smooth setting to define higher form symmetry algebras for Lie groups\footnote{This setting has been explored deeply in \cite{Schreiber}, 
and \cite{Grip} approaches generalized symmetries, 
with useful and clarifying discussions of examples and the subtleties of Postnikov towers.
Here I merely suggest that the compactly-supported version of their results will be useful.
Note that on a closed manifold, the global sections of our approach here should match with~\cite{Grip}.}
but see also Remark~\ref{rmk: better topos} below.\footnote{A more traditional approach might use instead  smooth variants of group cohomology, sometimes known as Segal-Mitchison cohomology, 
which are likely adequate for many purposes.
See, as a starting point, \cite{Bry, SegalCoh, WagWoc}.}

\begin{dfn}
The $\infty$-category of {\em smooth spaces} is $\Shv(\Mfld,\Spaces)$, i.e., sheaves of spaces over the site of smooth manifolds.
\end{dfn}

An ordinary smooth manifold determines such a smooth space, 
via the Yoneda embedding.
That is, a manifold $Y$ determines a sheaf $h_Y$ by $h_Y(X) = \Map^{\rm sm}(X,Y)$.

There are other important examples of manifest physical importance.
For $G$ a Lie group, there are smooth spaces 
\begin{itemize}
\item $BG$ that classifies smooth $G$-bundles,
\item $\Conn_G$ that classifies smooth $G$-bundles with connection,\footnote{At the level of a functor into simplicial sets, the $p$-simplices of $\Conn_G(M)$ are smooth $G$-bundles on $\RR^p \times M$ equipped with  connections along the fibers of the projection $\RR^p \times M \to \RR^p$. See the end of Section 2 in~\cite{Grip}} and
\item $\Flat_G$ that that classifies smooth $G$-bundles with {\em flat} connection.
\end{itemize} 
Taking compactly supported maps into these smooth spaces will offer a smooth version of important smooth 0-form symmetries.

When $A$ is an abelian Lie group,
then $BA$, $\Conn_A$, and $\Flat_A$ are themselves abelian group objects as smooth spaces,
so that one can iterate the process of taking the classifying space.
Thus, for instance, we have a smooth space $B^{q+1} A$, providing a smooth analog to the Eilenberg-Maclane space $K(A,q+1)$ we used in the topological setting.

\begin{rmk}
As an $\infty$-topos, smooth spaces admits a theory of Eilenberg-MacLane spaces and Postnikov towers.
See section 7.2.2 of~\cite{HTT} for generalities and \cite{Schreiber, Grip} for smooth spaces.
\end{rmk}

We want to talk about compactly supported smooth maps, rather than just smooth maps.
As a first step, we need to define  basepoints.

\begin{dfn}
A {\em pointed} smooth space is an object $\cY \in \Shv(\Mfld,\Spaces)$ equipped with a map $y: * \to \cY$ from the terminal object (the ``point'') in smooth spaces.
In other words, it is an object in the slice category~$\Shv(\Mfld,\Spaces)_{*/}$.
\end{dfn}

Given a basepoint~$y$, we can talk about compactly supported sections as follows.

Fix a smooth manifold $X$,
and let $K \subset X$ be compact in the usual topology on $X$ (not one of the $k$d-supportive topologies). 
Given a pointed smooth space $(\cY,y)$, let $\cY_c(K)$ denote the sections of $\cY$ with support in $K$, namely the fiber product in~$\Spaces$
\[
\begin{tikzcd}
\cY_c(K) \ar[r] \ar[d] & \cY(X) \ar[d] \\
* \ar[r,"y"] & \cY(X \setminus K)
\end{tikzcd}
\]
where the rightmost vertical arrow is the restriction map and where the bottom horizontal arrow views $y$ as the composite map
\[
X \setminus K \to * \xto{y} \cY,
\]
which is a constant map to the basepoint in~$Y$.

For an {\em open} subset $U \subset X$, the compactly supported sections $\cY_c(U)$ is the colimit of the diagram
\[
\cY_c(K_1) \to \cY_c(K_2) \to \cdots
\]
where 
\[
K_1 \subset K_2 \subset \cdots \subset U
\]
is a sequence of compact subsets whose union is~$U$.
In other words, a section with compact support in $U$ is a section with support inside one of these compact subsets $K_n$ of~$U$.
(Alternatively, one can take the colimit over all compact subsets of~$U$.)
This construction can be turned into a functor $\cY_c: \Mfld \to \Spaces$.

\begin{dfn}
Given a pointed smooth space $\cY$,\footnote{One might as well take $\cY$ to be connected.}
the {\em smooth higher group symmetry algebra} valued in~$\cY$ on a $d$-dimensional manifold~$X$
is the prefactorization algebra
sending $U \in \Open(X)$ to~$\cY_c(U)$.
\end{dfn}

Compactly supported maps into the smooth space $B^{q+1}A$ encodes a smooth $q$-form symmetry valued in $A$ (such as $A = U(1)$).
Compactly supported maps into the smooth spaces $BG$, $\Conn_G$, and $\Flat_G$ encode smooth 0-form symmetries.
More concretely, one can view these as encoding background fields of the following kind:
\begin{itemize}
\item $BG_c$ encodes a background principal $G$-bundle that is trivialized (or framed) outside some compact region,
\item $\Conn_{G,c}$ encodes a background principal $G$-bundle and connection (so a gauge field) that are trivialized outside some compact region, and
\item $\Flat_{G,c}$ encodes a background principal $G$-bundle with {\em flat} connection (so a curvature-zero gauge field) that are trivialized outside some compact region.
\end{itemize}
Physics discussions of generalized symmetries are often cast in terms of such background fields.

Inspired by Theorem~\ref{NAPD} for discrete generalized symmetries, we suggest the following.

\begin{expn}
These smooth higher group symmetry algebras satisfy a local-to-global property,
using a smooth version of the $k$-dimensionally supportive topology from Definition~\ref{dfn: ksupp}.\footnote{The proof of Theorem~\ref{NAPD} uses explicit manipulations with simplicial complexes that do not immediately admit smooth versions, so new ideas are needed for a proof. A proof strategy would also guide the search for the correct Grothendieck topologies.} 
\end{expn}

\begin{rmk}
\label{rmk: better topos}
We have discussed smooth spaces, but some other $\infty$-topoi might be better.
For example it is extremely useful to allow ``formal thickenings'' of manifolds, 
so that one can talk about Lie algebraic symmetries on an equal footing
and one can capture directly the usual Noether theorem,
as well as generalizations with dg Lie algebras and $L_\infty$ algebras.
To do this, one works with sheaves on a site bigger than $\Mfld$,
such as formal Cartesian spaces or, even better, on derived manifolds of some kind.
See \cite{Schreiber} as a starting place to learn about this ``synthetic differential geometry,'' some applications in physics, and its derived generalizations.
\end{rmk}

\subsection{Anomalies as factorized central extensions}
\label{sec: anomaly}

We offer here an analog to the idea of a central extension of a group,
which appears traditionally in physics when a group acts projectively on a physical system (e.g., on a Hilbert space).
In this analogy, the analog of a group is a prefactorization algebra valued in $\Top$, 
such as our higher group symmetry algebras $\Map_c(-,Y)$,
so we need to formulate central extensions in a way where there is a prefactorization analog.\footnote{This analogy was introduced by Beilinson and Drinfeld in \cite{BD}, 
and Kevin Costello proffered it as capturing `t Hooft anomalies in conversation.}

Recall that a central extension
\[
0 \to U(1) \to \widehat{G} \to G \to 1
\]
of a Lie group $G$ has an underlying principal $U(1)$-bundle: 
just forget the multiplication on $\widehat{G}$ and keep the projection map.
This fact suggests that a principal $U(1)$-bundle {\em and a little more data} encodes the whole central extension.

Consider, for instance, the multiplication map $\widehat{m}$ for $\widehat{G}$ from the bundle point of view.
Let $L \to G$ denote the underlying bundle of~$\widehat{G}$.
Let $m$ denote the multiplication map for~$G$.
There are two natural bundles over~$G \times G$,
\[
\begin{tikzcd}
L \times L \arrow[rd] & m^* L \arrow[r]\arrow[d] & L \arrow[d] \\
& G \times G \arrow[r, "m"] & G
\end{tikzcd}
\]
where $L \times L$ denotes the $U(1) \times U(1)$-bundle obtained by pulling back copies of $L$ along the projection maps.
It can be seen as a (non-principal) $U(1)$-bundle.
The map $\widehat{m}$ is then a bundle map~$L \times L \to m^* L$.

Similarly, the identity $\widehat{e}$ in $\widehat{G}$ is a point in the fiber $L_e$ over the identity $e$ in~$G$.
As $L_e$ is a $U(1)$ torsor, a distinguished point corresponds to a framing $U(1) \xto{\cong} L_e$ that sends $\lambda$ to~$\lambda \widehat{e}$.

A central extension corresponds to the following.

\begin{dfn}
A {\em multiplicative $U(1)$-bundle} on a Lie group $G$~is
\begin{enumerate}
\item a principal $U(1)$-bundle $L \to G$,
\item a framing $\phi: U(1) \cong L_e$ of the fiber over the identity~$e \in G$, and
\item a map of $U(1)$-bundles on~$G \times G$
\[
\mu: L \times L \to m^* L,
\] 
where $m: G \times G \to G$ is the multiplication
\end{enumerate}
satisfying natural coherence conditions.\footnote{For a complete definition see Chapter 10 of \cite{Pol} or, for the original treatment in the language of biextensions, see~\cite{Gro}.} 
\end{dfn}

A slick encoding is to say that a multiplicative $U(1)$-bundle is a map $c_L: G \to BU(1)$ of {\em group} stacks.
That is, $c_L$ is both a stack map and a group homomorphism, compatibly.

Viewing a prefactorization algebra $\cF$ valued in $\Top$ as analogous to a Lie group,
we can easily formulate an analog of a multiplicative bundle and hence of a central extension.

\begin{dfn}
Given a prefactorization algebra $\cF$ on $X$ valued in $\Top$,
a {\em factorizing $U(1)$-bundle} on $\cF$~is
\begin{enumerate}
\item a principal $U(1)$-bundle $L(V) \to \cF(V)$ for every open set $V \subset X$,
\item a framing $\phi(V): U(1) \cong L(V)_{e(V)}$ of the fiber over the basepoint~$e(V) \in \cF(V)$,\footnote{For every open $V$, the inclusion $\emptyset \hookrightarrow V$ determines a structure map $\cF(\emptyset) = \ast \to \cF(V)$ and hence a basepoint~$e(V)$.} and
\item for every finite collection $V_1,\ldots,V_k$ of pairwise disjoint open subsets of $W$,
a map of $U(1)$-bundles on~$\cF(V_1) \times \cdots \times \cF(V_k)$
\[
\mu^{V_1,\ldots,V_k}_W: L(V_1) \times \cdots \times L(V_k) \to m^* L(W),
\] 
where $m^{V_1,\ldots,V_k}_W: \cF(V_1) \times \cdots \times \cF(V_k) \to \cF(W)$ is the multiplication map of~$\cF$
\end{enumerate}
satisfying natural coherence conditions.\footnote{In the setting of algebraic geometry, Beilinson and Drinfeld offer a full definition in~\cite{BD}. An expository discussion appears in Chapter 20 of~\cite{FBZ}.}
\end{dfn}

Inasmuch as an anomalous symmetry of $G$ means that a central extension $\widehat{G}$ acts instead,
then an anomalous higher group symmetry means that a factorizing bundle acts.

It is natural to ask how to exhibit and construct any factorizing bundles,
or how to classify factorizing bundles.

As a partial answer, consider the special case of $\cF = \Map_c(-,Y)$ on the $n$-manifold $\RR^n$.
We saw earlier that it encodes the $E_n$ algebra structure of $\Omega^n Y$, the $n$-fold based loop space.
A factorizing $U(1)$-bundle $L$ on $\cF$ will determine, in particular, an $E_n$-algebra map
\[
c_L: \Omega^n Y \to BU(1)
\]
where we view $BU(1)$ as an $E_n$-algebra because it is, in fact, an $E_\infty$-algebra.\footnote{$BU(1)$ classifies complex line bundles, and it has a homotopy-coherent commutative product via tensoring of line bundles. 
Such an $E_\infty$ algebra structure can forget to an $E_n$-algebra structure.}
But
\[
\Hom_{\Alg_{E_n}}(\Omega^n Y, BU(1)) \simeq \Hom_{\Spaces}(Y, B^{n+1} U(1))
\]
by the $\Omega$-$B$ adjunction.
Thus $c_L$ encodes an $n+1$-cocycle valued in $U(1)$,
also known as an $n$-gerbe.
A heuristic argument using transgression indicates that one can use this gerbe to construct a factorizing bundle on any $n$-manifold (up to issues of orientation).

In short, a factorizing bundle on a higher group symmetry algebra valued in $Y$ is a higher gerbe on~$Y$.

This idea fits nicely with the role of anomalies in the functorial approach,
such as the twisting cocycles appearing in the $\pi$-finite TFTs that generalize Dijkgraaf-Witten theory.

\begin{rmk}
In the algebraic setting of Beilinson-Drinfeld factorization algebras,
there is a classification of factorizing line bundles on the Beilinson-Drinfeld Grassmannians that play a central role in the geometric Langlands correspondence~\cite{Gaits, TaoZhao}.
These results and their connections with WZW models suggest that the factorization point of view extends nicely beyond the TFT setting.
\end{rmk}

\printbibliography

\end{document}